\documentclass[a4paper,12pt]{article}
\usepackage{amsmath, amssymb, tensor, xcolor,graphicx, microtype, comment,soul}
\usepackage[hidelinks]{hyperref}
\newcommand{\be}{\begin{equation}}
\newcommand{\ee}{\end{equation}}

\newtheorem{theo}{Theorem}
\newtheorem{defi}{Definition}

\usepackage[style=ext-authoryear-comp,maxcitenames=2,uniquename=false,backend=biber,uniquelist=false,articlein=false]{biblatex}
\DeclareNameAlias{sortname}{family-given}

\title{How to Recover Spacetime Structure from Privileged Coordinates}

\date{20 August, 2024}

\author{Henrique Gomes,\footnote{gomes.ha@gmail.com}~ Tushar Menon,\footnote{tm399.acu@pm.me}\\Oliver Pooley,\footnote{oliver.pooley@philosophy.ox.ac.uk}~ \& James Read\footnote{james.read@philosophy.ox.ac.uk}}

\begin{document}

\maketitle

\begin{abstract}
    We show that the geometric structure of an arbitrary relativistic spacetime can be determined by the transformation groups associated with a collection of privileged coordinate systems.
\end{abstract}

\tableofcontents

\section{Introduction}\label{sec:intro}

Can the geometric structure of an arbitrary relativistic spacetime be presented via a collection of privileged coordinate systems? Recently, \textcite{BarrettManchak} have argued that there are non-isometric relativistic spacetimes that admit the same privileged coordinates. They take this to show that privileged coordinates do not always fully disclose spacetime structure. %
But their conclusion rests on a notion of privileged coordinates that is at once too narrow and too permissive. It is too narrow in the sense that they work with an unduly restrictive set of procedures to take one from privileged coordinate systems to spacetime structure. It is too permissive in the sense that, for a given structured space, \textit{any} coordinate system on that space can count as privileged, so long as one makes appropriate choices elsewhere.%

If one is presented with a set of privileged coordinates on a space, and is given no further information, how might one hope to recover the geometric structure of that space? The option that Barrett and Manchak focus on is recovery via a version of Klein's Erlangen programme. According to this point of view, the geometric structure of a space $S$ is associated with a privileged \emph{transformation group}, i.e., a privileged group of bijections from that space to itself. The geometry is identified as exactly the structure that is left invariant by the group. A natural generalisation replaces the requirement of invariance under a transformation group with invariance under a pseudogroup.%
\footnote{For details of this generalisation, see \textcite[\S3]{wallace-coord} and \textcite[\S\S2.2 and 3.2]{BarrettManchak}.}

What is the connection between such a group-theoretic characterisation of a geometry and preferred coordinate systems? We consider this question in more detail in the next section. For now, we simply note that coordinate systems can serve as a way of singling out the privileged transformation group (or pseudogroup) on $S$. Let $\phi \colon S \to \mathbb{R}^n$ and $\psi \colon S \to \mathbb{R}^n$ be two global coordinate charts on $S$.\footnote{We take a \emph{coordinate system} on $S$, in the most general sense of that term, to be any method of assigning a unique label to each point of $S$. It is a one-one function that maps a point to its label. A \emph{global} coordinate system on $S$ is simply a coordinate system whose domain is $S$. We use \emph{chart} to denote a coordinate system that encodes the topological and differential structure of $S$. (The relevant notion of encoding is discussed in the next section.) In this article, the coordinate systems which we use are also charts, and so we generally deploy the terms interchangeably.} Let's further stipulate that the ranges of $\phi$ and $\psi$ coincide. This means that the two charts define a bijection on $S$, namely, $\psi^{-1} \circ \phi$. More specifically, we can stipulate that the range of both charts is $\mathbb{R}^n$ itself, i.e., that the charts are bijections from $S$ to $\mathbb{R}^n$. \textit{Any} collection of global charts on $S$ meeting this condition defines a unique transformation group on $S$.\footnote{%
An arbitrary set $C$ of (not necessarily global) charts on a differentiable manifold $S$ will similarly define a subset $\Gamma_0$ of the diffeomorphism pseudogroup on $S$. Let $\text{Dom}(\phi)$ stand for the domain of $\phi$ and $\text{Ran}(\phi)$ stand for its range. $\Gamma_0$ is then defined as the set $\{\psi^{-1} \circ \phi : \phi, \psi \in C \}$ where $\text{Dom}(\psi^{-1} \circ \phi)$ is $\phi^{-1}[\text{Ran}(\phi) \cap  \text{Ran}(\psi)]$. While $\Gamma_0$ automatically satisfies the pseudogroup condition that \textcite[128]{wallace-coord} calls ``Closure'' and that corresponds to axioms PG5 and PG6 in \parencite{BarrettManchak}, it need not satisfy the other pseudogroup axioms. Nevertheless, it determines a pseudogroup since there is a unique minimal extension of $\Gamma_0$ that satisfies both Barrett and Manchak's axiom PG2 and, crucially, the condition that \textcite[128]{wallace-coord} calls ``Local definedness'' (cf.\ Barrett and Manchak's axiom PG3).%
}

If the role of a privileged set of charts on $S$ is merely to pick out a group (or pseudogroup) on $S$, then the question of whether particular geometric structure can be presented via such a set breaks down into two questions: (1) Can the geometry be characterised in terms of a group (or pseudogroup)? (2) Can the relevant group (or pseudogroup) be singled out via a privileged set of charts?

Presenting geometric structure in this way  (i.e., via privileged charts whose only role is to pick out a group which in turn picks out the structure) corresponds very closely to what \textcite{BM3} call a \textit{Kleinian presentation} of the structure. We will refer to it as a \textit{BMK presentation} of the structure, partly because our characterisation does not exactly follow Barrett and Manchak's (the difference is explored in the next section), and partly because we do not wish to suggest that this is in fact the correct `Kleinian' understanding of the role of privileged coordinates.

As we review in the next section, because any transformation group on $S$ can be encoded via a set of privileged charts, the answer to question (2) is ``yes.'' The question of whether geometry can be given a BMK presentation thus reduces to question (1).\footnote{%
Theorem 4.2.1 in \textcite[\S4.2]{BarrettManchak} is, effectively, a statement of this equivalence for the special case of relativistic spacetimes.%
}

In the Kleinian tradition, versions of question (1) have been extensively explored, quite independently of coordinate charts. Various classes of highly symmetric spaces are known to have group-theoretic characterisations. One example comprises the spaces (``Klein geometries'') that, for some Lie group $G$ and closed subgroup $H$, can be identified with the space $X = G/H$.\footnote{This case covers %
Minkowski spacetime, where $G$ is the Poincar\'{e} group and $H$ the Lorentz group; $n$-dimensional de~Sitter spacetime, where $G/H = O(1,n)/O(1,n-1)$ and $n$-dimensional anti-de~Sitter spacetime, where $G/H = O(2,n-1)/O(1,n-1)$.} In such a case $G$ acts as the automorphism group of the structure on $X$. Since it acts transitively, the spaces are \textit{homogeneous}: they `look the same' everywhere. The generalisation to pseudogroups that act transitively yields \textit{locally} homogeneous spaces: spaces that have a basis of open sets each member of which is isomorphic to an open set in some Klein geometry.%

Groups that do not act transitively on a space can also be discriminating in the structure that they determine. An example that will play a crucial role for us later is the structure of an $n$-dimensional real vector space $V$, which can be defined in terms of its automorphism group, $GL(n,\mathbb{R})$ (which does act transitively on $V \setminus \{0\}$). Additional structure on a vector space, such as an inner product, can be characterised by selecting an appropriate proper subgroup as the automorphism group.\footnote{Vector space structure is a central recurring example in \parencite[\S2]{wallace-coord}.}

All of these cases involve the characterisation of a highly symmetric structured space in terms of its (correspondingly highly non-trivial) symmetry group. Unsurprisingly, a structured space that \textit{lacks} symmetries cannot be characterised in terms of its symmetry group and therefore cannot be given a BMK presentation. The crucial observation is that, for many types of structure, there are non-isomorphic structures of that type that lack any symmetry. Such structures cannot be distinguished in terms of their symmetry groups since their symmetry groups are one and the same: the group whose sole element is the identity map.

Let us specialise to the case where the structure in question can be represented by a tensor field $\alpha$ on a smooth manifold $M$. In this case, a diffeomorphism $d$ will be a  symmetry of the structure just in case $d^*\alpha=\alpha$. The group of such diffeomorphisms is the automorphism group of the structure. And one can say that the structure lacks any symmetry just in case the only such diffeomorphism is the identity map.\footnote{More carefully, one should consider the sub-pseudogroup of the diffeomorphism pseudogroup defined by the condition $d^*\alpha=\alpha$, now with $d$ a diffeomorphism between open subsets of $M$. The automorphism group of a structure without symmetries will be the pseudogroup whose elements are all and only the identity maps on each open subset of $M$.} In this case the crucial fact is that, for two tensor fields $\alpha, \beta$ of a given type (e.g., $(0,2)$-tensor fields symmetric in their action on vector fields), there are many cases where, for $d$ a diffeomorphism, (i) $d^*\alpha=\alpha$ only if $d = \text{Id}$; (ii) $d^*\beta=\beta$ only if $d = \text{Id}$; but (iii) there is no $d$ such that $d^*\beta=\alpha$.\footnote{We take this fact to be well known. The sceptical reader is invited to consider the proof of Proposition~3.2.4 in \parencite{BarrettManchak}.} Such structures are therefore not characterised by their automorphism groups and so \textit{a fortiori} cannot be given a BMK presentation.

We opened this section with the question whether the geometric structure of an arbitrary relativistic spacetime can be presented via a collection of privileged coordinate charts. Barrett and Manchak's principal punchline is that, \textit{if the presentation is to be a BMK presentation}, the answer is ``no'' \parencite[see, in particular,][Theorem 3.2.1]{BarrettManchak}. Despite the fifteen pages of meticulous setup through which Barrett and Manchak take their reader in order to get to this point, the result should not have come as a surprise. Characterising a spacetime geometry via its symmetry group is simply a non-starter when that spacetime lacks symmetries.

It is worth stressing that spacetimes lacking symmetries are well-known to be the generic case. Barrett and Manchak cautiously conjecture that ``spacetimes determined by local isometry are much more the exception than the rule'' \parencite*[17]{BarrettManchak}. This is obviously correct. In \textcite[\S3.2]{BM3} they consider the condition of determination \textit{up to homothety} by local isometry and prove that at least every flat relativistic spacetime satisfies this weaker condition. It is clear that spacetimes satisfying this condition are also very much the exception. Here is an intuitive way of seeing this: an isometry group is a finite-dimensional closed subgroup of the infinite-dimensional group of diffeomorphisms, and a non-trivial isometry implies the components of the metric are the same along the orbit of that isometry. But components of the metric can be relatively arbitrary smooth functions. Even if one imposes the Einstein field equations, as a field the metric has two physical (i.e.\ unconstrained) degrees of freedom \textit{per spacetime point}. Therefore, starting from a metric with some isometry group, one could intuitively consider infinitesimal disturbances of the metric at any point along the isometry's orbit, which would therefore spoil it.   Indeed, we could trace back to \textcite[p.~1001]{Ebin1968} a proof, valid in the case of Riemannian signature, that metrics with trivial isometries are generic (open and dense) in any of the natural topologies of the space of metrics. But we suspect this fact has a much older lineage; we would not be surprised if it can be traced back to Riemann's writings. Today a lot more is known about the structure of the space of metrics. For instance, in the case of a closed spatial topology, \textcite{Fischer1970} 
showed that metrics with isometry groups form stratified manifolds, with those with more symmetry lying at the boundary of those with less symmetry.

Although one cannot disclose generic relativistic spacetime structure via a BMK presentation, it can be done via a natural generalisation of the procedure. The structure of a particular Lorentzian manifold can be thought of as a (point-by-point) deformation of Minkowski space, just as a Riemannian manifold can be thought of as a (point-by-point) deformation of Euclidean space. In this context, sets of privileged coordinates are naturally indexed to particular points. The relevant question becomes whether one can recover the geometric structure of the manifold given a privileged set of coordinates \emph{for each point of the manifold} (and given no other information!).

In \S\ref{recovering}, we describe how this can be done. While the recovery route is very close in spirit to the one involved in a BMK presentation, it involves a crucial generalisation. The privileged set of coordinates at a point $p$ will single out a transformation group not on a neighbourhood of $p$ itself but on certain spaces associated with $p$ (such as its tangent space).\footnote{\label{fn-cartan}A different way of generalising Kleinian geometry to inhomogeneous spaces yields Cartan geometries \parencite[see, e.g.,][]{wiseMacDowellMansouriGravity2010}. Although philosophically interesting, we won't discuss Cartan geometries further in this article.}
Nevertheless, the essential core remains: privileged structure associated with a point $p$ is identified via the condition that the structure is left invariant by a transformation group. The geometrical structure of the manifold as a whole can then be thought of as simply the sum total of these point-relative structures taken together.

In \S\ref{recovering} we review the details, and in \S\ref{close} we survey the prospects for further philosophical application of this work. Before that, some further stage-setting is appropriate. In \S\ref{pc}, we clarify what it means for a set of coordinates to be privileged. In \S\ref{ac}, we review various types of privileged coordinate systems for relativistic spacetimes.

\section{Privileged Coordinates}\label{pc}

The label `privileged coordinate systems' is ambiguous. Does the privilege attach, in the first instance, to the collection as a whole, or does the privilege attach to each chart individually, with their intrinsic privilege determining their membership in the collection?

In the previous section we were careful only to talk about privileged \emph{sets} of charts. We described how a set of global charts on a space is associated with a privileged transformation group on that space. It is crucial to note that a chart's membership in such a set does not place any constraints on the chart itself. Suppose that we wish to define a set of charts corresponding to a given transformation group $\Gamma$ on space $S$. Let $\phi$ be an arbitrary diffeomorphism from $S$ to $\mathbb{R}^n$. In order for a set $C$ containing $\phi$ to define $\Gamma$, we just need to choose the right co-members. We simply stipulate that $\phi' \in C$ if and only if $\phi' = \phi \circ \gamma$ for some $\gamma \in \Gamma$.\footnote{%
Suppose that $\Gamma$ is instead a pseudogroup on $S$. There will be similarly arbitrary membership conditions for any set $C$ of local charts on $S$ that encodes $\Gamma$ via suitable constraints on $\psi^{-1} \circ \phi$ for $\phi, \psi \in C$. In particular, for $d$ an arbitrary diffeomorphism on the value space of the charts, the set $\{d \circ \phi: \phi \in C \}$ will encode $\Gamma$ in exactly the same way as $C$. For one way to associate a pseudogroup with a set of charts, see \textcite[Definition 2.2.2]{BarrettManchak}.}

Let us call a chart considered as belonging to such a set a \textit{symmetry-specifying chart}. 
Barrett and Manchak treat symmetry-specifying charts as privileged but such charts are typically privileged only in a highly etiolated sense. First, a symmetry-specifying chart taken by itself tells us nothing about the group that it plays a role in defining. One needs two such charts to identify a group element (other than the identity) and it is only the collection of charts taken together that defines the group. Second, as we have just seen, for any given target group any chart can be considered a symmetry-specifying chart.

One therefore suspects that Barrett and Manchak's conception of a privileged chart does not correspond to what is normally meant by this term and, indeed, a better candidate is ready to hand. A chart might be said to be privileged because it is `adapted to' or `encodes' geometric structure. We will call charts that are privileged in this sense \emph{adapted charts}. 
The idea is best illustrated by example.

Consider the simple case of the Euclidean plane, $\mathbb{E}^2$. A \textit{Cartesian coordinate system} on $\mathbb{E}^2$ is defined in terms of, and thus adapted to, 
the geometric structure of $\mathbb{E}^2$. The coordinate axes are chosen to be \textit{straight}, \textit{orthogonal} lines, and the coordinate values of a point are given by the (signed) \textit{distances} from the corresponding axis. The terms emphasized in the previous sentence all advert to the geometric structure of $\mathbb{E}^2$. The definition secures that the distance $d(p,q)$ between two points $p$ and $q$ with coordinates $(x_p,y_p)$ and $(x_q, y_q)$ is encoded by the simple coordinate function $d(p,q) = \sqrt{(x_p - x_q)^2 + (y_p - y_q)^2}$. Equivalently, one can think $\mathbb{R}^2$ itself as a structured space and think of $\sqrt{(x_p - x_q)^2 + (y_p - y_q)^2}$ as the canonical metric on $\mathbb{R}^2$. A Cartesian coordinate system is then just an isometry from $\mathbb{E}^2$ to $\mathbb{R}^2$. More generally, adapted coordinate systems can be thought of as structure preserving maps from the structured space $S$ to $\mathbb{R}^n$.

Such structure preserving maps normally will not be unique and it is this non-uniqueness that provides the more usual connection between privileged charts and a Kleinian definition of geometric structure. In our example, there are many structure preserving maps from $\mathbb{E}^2$ to $\mathbb{R}^2$ understood as equipped with its canonical metric. Such charts will be related by exactly the symmetry group of the Euclidean plane%
. They therefore also constitute a collection of symmetry-specifying charts.

When first presenting their formal machinery, \textcite[\S2.1]{BarrettManchak} relate a set of charts (their set $C$) to a transformation group on $S$ (their ``coordinate transformation group'', $\Gamma$). They do not, however, (as we did above) characterise $C$ in terms of $\Gamma$. Rather, their direction of definition is reversed. $\Gamma$ is defined in terms of $C$ which in turn is first characterised group-theoretically (but non-uniquely), via a transformation group $G$ on $\mathbb{R}^n$. Their compatibility condition on the members of $C$ states: ``if $f \in C$, then $f' \in C$ if and only if $f \circ f'^{-1} \in G$'' \parencite*[p.~4]{BarrettManchak}. \textcite[127]{wallace-coord} also characterises the sets of coordinate systems to be associated with structure on the coordinatized space in terms of transformation groups on $\mathbb{R}^n$.

Armed with the notion of adapted charts, we are in a position to see why prioritising a group action on $\mathbb{R}^n$ is well-motivated from Wallace's perspective but not from Barrett and Manchak's. A bijection between spaces induces a natural map that carries structure from one space to the other.\footnote{If the bijection is a diffeomorphism, the induced map acting on geometric objects is the pushforward map but this is just a specific example of a much more general concept.} The map induced by an adapted chart carries the structure on $S$ onto a privileged substructure of $\mathbb{R}^n$. Let $\Gamma$ be the transformation group on $S$ that preserves the structure of $S$. Let $G$ be the transformation group on $\mathbb{R}^n$ that preserves the privileged substructure of $\mathbb{R}^n$. %
Let $\phi$ be an adapted chart (i.e., a structure preserving map). It follows that $\phi \circ \gamma$ will be an adapted chart if and only if $\gamma \in \Gamma$ (so the set of adapted charts is always a set of symmetry-specifying charts), and that $g \circ \phi$ will be an adapted chart if and only if $g \in G$.

One can see $G$ as picking out the relevant substructure of $\mathbb{R}^n$ to be targeted by an adapted chart. $\mathbb{R}^n$ is a highly structured space but one is interested in only some of this structure, namely, just that structured preserved by the relevant transformation group on $\mathbb{R}^n$.\footnote{This is why \textcite{norton1999collision} describes Klein's approach as a ``subtractive strategy.''} This perspective allows one to view privileged charts not as adapted to antecedently-given structure but rather as \textit{defining} such structure on $S$. The structure to be defined is the pullback to $S$ of exactly that structure preserved by $G$. Call charts conceived of as privileged in this way \textit{structure-defining charts}. \textcite{wallace-coord} can be read as defending the propriety of this point of view. Structure-defining charts are, of course, automatically adapted to the structures they are used to define.

As noted, Barrett and Manchak follow Wallace in defining their privileged charts in terms of a transformation group (or pseudogroup) $G$ defined on $\mathbb{R}^n$. They further consciously adopt Wallace's terminology of ``$G$-structured spaces''. However, unlike Wallace's, their privileged charts need not be adapted charts. This is the case even if one views their charts as defining as structure on $S$ whatever structure is preserved by the coordinate transformation group $\Gamma$ defined by the charts. The charts need not be adapted because Barrett and Manchak's group $G$ is not required to be an automorphism group of some natural structure on $\mathbb{R}^n$.\footnote{While not required for what they call Kleinian presentability, some of their examples do involve adapted charts. See, in particular, the examples in \parencite[\S2]{BM3}.} 

It will be instructive to see how their more permissive notion of privileged charts---charts that are symmetry-specifying but that need not be adapted---are used in what they would call a Kleinian presentation of the geometry of the Euclidean plane. Barrett and Manchak stipulate that the transformation group $G$ on $\mathbb{R}^n$ to be associated with the geometry should be a subgroup of the diffeomorphism group of $\mathbb{R}^2$. We therefore restrict to diffeomorphisms from $\mathbb{E}^2$ to $\mathbb{R}^2$; let $f$ be one such bijection (and therefore a chart). Generically, in such a chart the components $g_{ij}$ of the metric tensor of $\mathbb{E}^2$ will be smooth but otherwise arbitrary functions of position, in principle varying wildly from point to point.  Coordinate intervals in $f$ bear no obvious relation to actual distances in $\mathbb{E}^2$, and there is no sense in which $f$ is adapted to the geometry of $\mathbb{E}^2$.

Nevertheless, we can use $f$ to define a non-standard metric $f_{*}g_{ab}$ on $\mathbb{R}^{2}$, where $g_{ab}$ is the metric tensor on $\mathbb{E}^2$.%
\footnote{Here we are considering $\mathbb{R}^{2}$ just as a differentiable manifold and ignoring the rest of its structure.}
We then define $G$ to be the subgroup of the diffeomorphism group of $\mathbb{R}^{2}$ that preserves this non-standard distance function on $\mathbb{R}^2$. ($G$ will be isomorphic as a group to $E^2$ but it will correspond to a non-standard realization of $E^2$ on $\mathbb{R}^2$.) Next, we consider the full set $C$ of coordinate systems on $\mathbb{E}^2$ related to $f$ via Barrett and Manchak's ``compatibility condition'': $f' \in C$ if and only if $f \circ f'^{-1} \in G$. Finally, we can use this set to define a transformation group $\Gamma$ on $\mathbb{E}^2$: $\Gamma = \{ c^{-1} \circ d : c, d \in C \}$.

By construction, $\Gamma$ is the standard Euclidean group on $\mathbb{E}^2$, but this fact is doubly obscure if all one is presented with is the set $C$. First, as has already been stressed, each individual chart in $C$ tells us nothing about the geometric structure of the space that it coordinatizes. All elements of $C$ `represent' the distance function on $\mathbb{E}^2$ in the same way: its component functions on $\mathbb{R}^2$ are identically the same in all such coordinate systems. But these functions will typically be a complicated mess exhibiting no identifiable regularity or pattern. Second, since $G$ is a non-standard realization of $E^2$ on $\mathbb{R}^2$, that $\Gamma$ is the Euclidean group will not be evident from simple inspection of the functional form of the transition functions $c \circ d^{-1} \in G$, $c,d \in C$. 

The somewhat involved nature of this construction prompts one to ask why \textcite{BarrettManchak} tacitly forgo adapted charts.\footnote{Things become even more cumbersome when generalising to pseudogroups and to a space $S$ that is not diffeomorphic to $\mathbb{R}^n$. In such a case one cannot simply push forward the structure of $S$ to $\mathbb{R}^n$. The reader is invited to acquaint themselves with the proofs of Lemmas 3.2.1 and 3.2.2 in \parencite{BarrettManchak} for Barrett and Manchak's solution to this problem, in terms of their notion of a `representation' of a spacetime on $\mathbb{R}^n$. The root cause of the unnaturalness stems from  starting with a group or pseudogroup on $\mathbb{R}^n$, despite eschewing adapted charts. Defining a condition on a set of charts directly in terms of the automorphism (pseudo)group on $S$ might avoid some of the difficulties.} When dealing with generic relativistic spacetime structure, one might have supposed that one has no choice. Given that the structure of a generic Lorentzian geometry lacks any symmetries, there will be no way to encode its structure in terms of simple relationships between coordinate values. To put the point another way, the structure of a generic Lorentzian geometry is not isomorphic to any `natural' structure definable on $\mathbb{R}^4$. If adapted charts are required to map the structure of the spacetime to natural structures on $\mathbb{R}^4$, generic Lorentzian geometries lack adapted charts.

In another sense, however, adapted charts still exist, so long as the adaptation is localised and \emph{relativised to a point}.\footnote{Since we will not need it, we set to one side the possibility of relativisation to submanifolds, such as that involved in, e.g., the definition of Fermi normal coordinates.} It is adapted charts of this kind that can be naturally thought of as a window (really a family of infinitesimal windows) on relativistic spacetime structure. The next section reviews some of the standard ways localised notions of adapted charts can be defined. We then consider how spacetime structure can be recovered from such sets of charts.

\section{Adapted Coordinates for Relativistic Spacetimes}\label{ac}

Let $(M,g)$ be a four-dimensional relativistic spacetime.\footnote{The restriction to four dimensions is simply for ease of exposition.} $M$'s global topology might preclude global charts. But for each $p \in M$, there will be an open neighbourhood $U_p$ of $p$ such that there is a diffeomorphism $\phi$ from $U_p$ into $\mathbb{R}^4$. Such a diffeomorphism is a chart for $U_p$.

We can think of $\phi$ as a family of four \textit{coordinate functions}: four smooth real scalar functions, 
$x^{\mu}$, $\mu = 0,1,2,3$, defined on $U_p$; $\phi$ maps $q$ in $U_p$ to $(x^0(q),x^1(q),x^2(q),x^3(q))$ in $\mathbb{R}^4$. %
In these terms, we can define two notions that will be useful in what follows.
\begin{description}
    \item[Level surfaces of a coordinate:] A level surface of a coordinate function defined on $U_p$ is a maximal set of points in $U_p$ that are all assigned the same value by that function. The level surfaces of a given coordinate partition $U_p$ into 3-dimensional submanifolds.\footnote{This follows because $dx^0\wedge dx^1\wedge dx^2\wedge dx^3\neq 0$ implies, for any subset of three coordinates, that $dx^i\wedge dx^j\wedge dx^k\neq 0$.} The elements of the partition can be parameterised by the value of the relevant coordinate on each surface.
    \item[Coordinate curves:] The intersection of three such surfaces, defined by particular values of three of the four coordinates, is a 1-dimensional submanifold of $U_p$. We can think of this manifold as the image of a curve parameterized by the fourth, non-constant coordinate. Again, the family of curves associated with a given coordinate partitions $U_p$. To take a concrete example, consider the coordinate curves associated with $x^0$. The various possible constant (but in general differing) values of $x^i$, $i=1,2,3$ parameterise the elements of the partition. The parameter value of any point $q$ on any given curve is just $x^0(q)$.
\end{description}

We now proceed to characterise seven distinct notions of adapted coordinate system and to spell out the logical relationships between them. In what follows $p$ is an arbitrary point of $M$. 

\subsection{Local Lorentz and Local Conformal Charts}

In the previous section, Cartesian coordinates on $\mathbb{E}^2$ were the paradigm example of adapted coordinates.  %
They diagonalise the Euclidean metric and set the values of its diagonal elements to 1. Similarly, in \emph{Lorentz charts} on Minkowski spacetime, the metric everywhere takes the canonical form $\text{diag}(-1,1,1,1)$. In a non-flat Lorentzian spacetime, we cannot choose coordinate systems such that the metric $g_{ab}$ everywhere takes this form. But we can always choose a coordinate system such that this condition holds at a point. Call a chart $\{x^{\mu}\}$ defined on $U_p$ a \emph{local Lorentz chart for $p$} iff (i) $x^{\mu}(p) = 0$ ($\mu = 0,1,2,3)$ and (ii), \emph{at $p$}, $g_{\mu \nu} = \eta_{\mu \nu}$.\footnote{We use $\eta_{\mu \nu}$ to refer to the $\mu \nu$-components of the numerical matrix $\text{diag}(-1,1,1,1)$. Note that the role of (i) is simply to fix a common origin for the class of coordinate systems. It plays no role in characterizing properties of the metric. In the case of normal coordinates (described below), (i) also holds but is not an independent stipulation.}\textsuperscript{,}\footnote{Ehlers calls such coordinates ``locally inertial'' at $p$ \parencite[94]{ehlers07gen}. Since `inertial' suggests a connection to force-free motions, and hence to geodesic structure, we prefer a different label.} In other words, local Lorentz charts for $p$ diagonalise the metric at $p$ (as well as implement the convention that, at $p$, $x^0$ is the time coordinate).

The metric at a point defines the relative lengths of (and (pseudo-)angles between) tangent vectors at that point but it also defines a notion of `absolute' length (which can be understood as simply encoding a path-independent notion of the relative lengths of vectors at different points). %
This notion of absolute length is encoded in local Lorentz charts by the condition that the absolute value of the diagonal elements of the metric is unity. Relaxing this condition, and just requiring that (at $p$) $g_{\mu \nu} = \lambda \eta_{\mu \nu}$ for some $\lambda \in \mathbb{R}^+$, yields the \emph{local conformal charts} (for $p$).

\subsection{Locally Affine and Locally Projective Charts}\label{affine}

Call a chart \emph{locally affine at $p$} iff (i) $x^{\mu}(p) = 0$ and (ii) the connection coefficients, $\Gamma^{\mu}_{\phantom{\mu}\nu\lambda}$, vanish at $p$.\footnote{\label{fn-normal}There is a minority, although established, practice of labelling these charts ``normal coordinates''. We follow more standard practice in reserving this label for the proper subclass of locally affine charts described in \S\ref{s:normal}.} Geometrically, this requires that the chart be adapted to the geodesic structure of $(M,g)$ in the `infinitesimal neighbourhood' of $p$. More specifically, a necessary and sufficient condition for the connection coefficients to vanish is that each coordinate curve is (a) `geodesic at $p$' (i.e., both straight and affinely parameterized at $p$) and (b) parallel to all the curves corresponding to the same coordinate in the infinitesimal neighbourhood of $p$, which should also be affinely parameterized in that neighbourhood. Let $\partial_{\mu}$ be the vector field corresponding to the coordinate $x^{\mu}$, i.e., the $x^{\mu}$ coordinate curves are integral curves of $\partial_{\mu}$. The two conditions then correspond to the condition $\nabla_{\partial_{\nu}}\partial_{\mu} = 0$ at $p$, with (a) corresponding to the case $\nu = \mu$ and (b) corresponding to the case $\nu \neq \mu$. But since $\nabla_{\partial_{\nu}}\partial_{\mu} =: \Gamma^\rho_{\phantom{\rho}\mu\nu}\partial_{\rho}$, this condition is just what is required for all the connection coefficients to vanish.

If one relaxes the condition that the coordinate curves should be affinely parameterized, one obtains the \emph{locally projective charts at $p$} and the corresponding condition on the covariant derivatives of the coordinate curves at $p$ becomes $\nabla_{\partial_{\nu}}\partial_{\mu} = \lambda \partial_{\mu}$.

\subsection{Lorentz Affine Charts}

Whilst the condition of being a local Lorentz chart is strictly stronger than the condition of being a local conformal chart, and whilst being locally affine is strictly stronger than being locally projective, these two pairs of conditions are logically independent from one another. A chart might be a local Lorentz chart for $p$ without being locally affine at $p$. And a chart might be locally affine at $p$ without being a local Lorentz chart for $p$. This observation leads to our next notion of adapted coordinates: we call charts that satisfy both conditions \emph{(locally) Lorentz affine charts for $p$}.\footnote{These charts are sometimes called Lorentz (or even Riemann) normal coordinates. See, e.g., \textcite[7]{fletcherweatherall23matter} and \textcite[p.~20]{BarrettManchak}. We reserve this label for the strictly stronger condition described in \S\ref{LNCs}. Cf.\ footnote~\ref{fn-normal}.}

An alternative characterisation of Lorentz affine charts is as ones in which (i) $x^{\mu}(p) = 0$, (ii) the metric is diagonalised at $p$, and (iii) the first derivatives of the metric vanish at $p$. The equivalence of the third condition with the condition that the connection coefficients vanish follows from the connection in question being the unique (torsion-free, symmetric) compatible connection. Given that $\nabla_{\rho} g_{\mu \nu} = 0$, $g_{\mu\nu,\rho} = 0$ iff $\Gamma^{\alpha}_{\phantom{\alpha}\mu\rho} = \Gamma^{\alpha}_{\phantom{\alpha}\nu\rho} = 0$.

\subsection{Normal Charts}\label{s:normal}

Our next two notions of adapted charts are much more restrictive than the ones so far surveyed. They use the \emph{exponential map}, a coordinate-independent (but connection-dependent) diffeomorphism from a neighbourhood $U_0$ of the zero-vector of the tangent space $T_pM$ at $p$, to a neighbourhood $U_p$ of $p$%
.
The map is used to project a coordinate system on $T_pM$ (given by some set of basis vectors) onto $U_p$.

The exponential map associates each geodesic through $p$ with the one-dimensional linear subspace of $T_pM$ that contains the tangent vector to that geodesic. Let $\lambda(s)$ be a specific geodesic through $p$ with affine parameter $s$ such that $\lambda(0) = p$. We denote its tangent vector at $p$ as $\dot{\lambda}_p$. The exponential map is then defined as the map that takes $s\dot{\lambda}_p \in T_pM$ to $\lambda(s) \in M$. In particular, $\dot{\lambda}_p$ is mapped to $\lambda(1)$. When every point in $U_p$, the image of $U_0 \subset T_pM$ under the exponential map, is connected to $p$ by exactly one geodesic, the map is a bijection (and in fact a diffeomorphism) from $U_0$ to $U_p$. $U_p$ is then called a \textit{normal neighbourhood} of $p$.\footnote{This is a slight abuse of terminology, since `normal neighbourhood' refers in the first instance to $U_0$ rather than to its image under the exponential map \parencite[see, e.g.,][]{sachs2012general}.}

Given any basis $\{\mathbf{e}_0,\mathbf{e}_1,\mathbf{e}_2,\mathbf{e}_3\}$ for the tangent space $T_pM$ at $p$ we can use the exponential map to define a chart for any normal neighbourhood $U_p$ related to that basis. We simply assign to a point in $U_p$ the components of the vector that the exponential map sends to that point. In other words, if $s\dot{\lambda}_p = X^a\mathbf{e}_a$, the corresponding chart maps $\lambda(s)$ to $\{ X^0,X^1,X^2,X^3\}$. Such coordinate systems constitute all and only the \emph{normal charts for $p$}.

It is straightforward to establish that normal charts are locally affine charts, i.e., that the connection coefficients vanish at their origin. But they are a very special subclass of locally affine charts. The geodesics through $p$ have a very simple coordinate expression everywhere in the domain of any such chart, namely as $x^{\mu}(s) = sa^{\mu}$, where each $a^{\mu}$ is a constant. The coordinate transformation between any two such charts is linear, i.e., the coordinate transformation group is $GL(4)$.\footnote{\label{fn-gl4}More precisely, for a given point $p \in M$, the transition functions between its normal charts result from restricting $GL(4)$ to the ranges of the charts. (Note that, while two normal charts for $p$ need not share a domain, the intersection of their domains is always the domain of another normal chart.) The union of the all the ranges of the charts is fixed by the action of $GL(4)$ on $\mathbb{R}^4$. Note that the collection of all the transition functions  between normal charts for the given point $p$ does not form a pseudogroup on this subspace of $\mathbb{R}^4$ because, e.g., the domains of the transition functions must always include the origin of $\mathbb{R}^4$.}

\subsection{Lorentz Normal Charts}\label{LNCs}

Just as the conditions of being a local Lorentz chart for $p$ and being locally affine at $p$ can be combined, to give us the notion of a Lorentz affine chart, so one can supplement the condition of being normal at $p$ with the condition of being a Lorentz chart at $p$. This gives us the set of \emph{Lorentz normal charts for $p$}. They correspond exactly to the coordinate systems that are definable using the exponential map when one restricts to orthonormal bases for $T_pM$ (with, by convention, $\mathbf{e}_0$ as the timelike basis vector).

Since Lorentz normal charts are a proper subset of normal coordinates, the coordinate transformations will be a proper subgroup of $GL(4)$. Lorentz normal charts are related by exactly Lorentz transformations.\footnote{\label{fn-lnc-tf}More precisely, their transition functions are the result of restricting $O(1,3)$ on $\mathbb{R}^4$ to the ranges of all such charts for the given point $p$ (cf.\ the previous footnote). For further recent philosophical discussion of these normal coordinate constructions via the exponential map, see \parencite{TehForthcoming-TEHTLV}.}

\subsection{Summary}

We have surveyed seven distinct types of adapted coordinates systems definable in relativistic spacetimes. All are centred on a particular (arbitrary) point.\footnote{Not all adapted charts for a variably curved spacetime need be centred on a particular point. Examples that do not privilege a particular point include \emph{Gaussian normal coordinates} \parencite[see, e.g.,][42]{wald2010general} and what one might simply call \emph{spacetime charts}: coordinates where the level surfaces of $x^0$ are everywhere spacelike and where all of the $x^0$ coordinate curves are everywhere timelike.} Five of these---local conformal charts, local Lorentz charts, locally projective charts, locally affine charts, and local Lorentz affine charts---involved adaptation of the chart to the geometric structures only in the infinitesimal neighbourhood of that point: a given chart's meeting the condition imposes no constraints on the coordinate values assigned to any point other than $p$ in $U_p$.\footnote{All the charts that we consider are, of course, adapted in the same way to the local differentiable structure at each point in the chart's domain. This follows from our initial choice to consider only bijections from $U_p$ to $\mathbb{R}^4$ that are diffeomorphisms.} Intuitively, given any chart meeting the relevant condition, one can mess around arbitrarily with the chart outside of any finite neighbourhood of the point and the result will still meet the condition. More specifically, suppose $\phi$ is a chart defined on $U_p$ meeting one of the conditions at $p$. Let $d$ be a diffeomorphism on $\mathbb{R}^4$ and let $\phi'$ be the chart defined via $\phi' = d \circ \phi$. $\phi'$ will also meet the condition if (but not only if) 
$d$ reduces to the identity on a neighbourhood of $(0,0,0,0)$. That neighbourhood can be as small as one likes. So, for any $q \neq p$ in $U_p$, one can choose a neighbourhood of $(0,0,0,0)$ that does not contain $\phi(q)$. And for any $x \neq (0,0,0,0)$ in $\mathbb{R}^4$ one can choose a $d$ that preserves the coordinate condition but that sends $\phi(q)$ to $x$.

The two types of coordinate system based on the exponential map were adapted to certain geometric structures not definable purely in the infinitesimal neighbourhood of $p$ (which curves at $q$ are geodesic at $q$ depends on the geometric structure in the infinitesimal neighbourhood of $q$, and which of those curves is a geodesic curve connecting $q$ to $p$ depends on the geometric structure around every point on the curve between $q$ and $p$). Nevertheless, they still privilege geometric structure at $p$.

\begin{figure}
    \centering
    \includegraphics[width=0.5\linewidth]{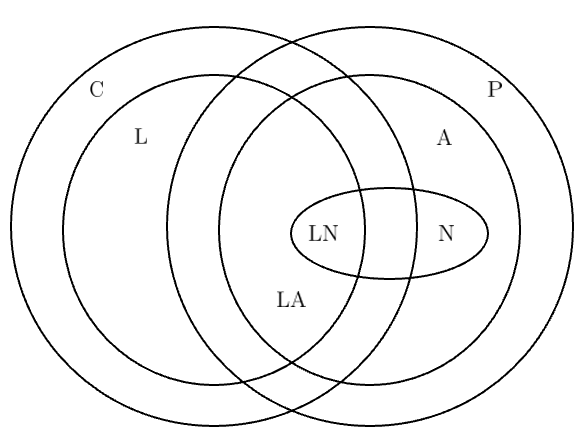}
    \caption{The logical relationships between: (C) conformal charts; (L) Lorentz charts; (P) projective charts; (A) affine charts; (LA) Lorentz affine charts; (N) normal charts; and (LN) Lorentz normal charts.}
    \label{fig:venn}
\end{figure}

Some of the logical relationships between these types of coordinate system are summarised in figure~\ref{fig:venn}. We now turn to the task of reconstructing geometric structure from such charts.

\section{Recovering Structure from Local Lorentz Charts}\label{recovering}

As we saw in \S\ref{sec:intro}, privileged charts understood as symmetry-specifying charts might fail to provide a BMK presentation of the geometry of a Lorentzian spacetime. This is because, while one can always find charts that determine the geometry's isometry pseudogroup, that pseudogroup might fail to determine the geometry. Having fastidiously established this result, \textcite[§5]{BarrettManchak} pose the question whether there might be a better choice of privileged charts from which to seek to recover the geometry. One option that they consider is given by (what we have labelled) the Lorentz affine charts.\footnote{As noted in footnote \ref{fn-normal}, they refer to these as Lorentz normal charts.} Their assessment is that if this is one's choice of privileged charts then ``one is forced to dramatically change the procedure with which one recovers the structure of $(M, g_{ab})$'' \parencite*[p.~20]{BarrettManchak}.

In a subsequent paper, they consider the recovery of the geometric structure of a spacetime $(M,g)$ from its Lorentz affine charts in a little more detail \parencite[\S4.2]{BM3}. They first consider applying their version of the Kleinian method by using such charts to define a pseudogroup on $M$. Unsurprisingly this fails, since the Lorentz affine charts were not designed to be symmetry-specifying charts in the sense of \S\ref{pc}. The pseudogroup that they define is not the isometry pseudogroup of the metric to which they are adapted.

Barrett and Manchak then consider a route by which one can successfully recover $(M,g)$ from its Lorentz affine charts: for each $p \in M$ pick an arbitrary Lorentz affine chart for $p$. Use that chart alone to explicitly define the metric at $p$ (we review this process in more detail below). Doing this for every $p$ defines $g$.

Their verdict on this procedure is that ``it does not represent a victory for proponents of Kleinian methods.'' This is because they hold that:
\begin{quote}
Kleinian methods are distinctive because they employ a variety of implicit deﬁnability, looking to those structures that are `invariant under symmetry'. %
[\ldots] The method of presentation suggested [\ldots] %
is not Kleinian in this sense. \parencite*[16]{BM3}
\end{quote}
We agree that, in order to merit the label `Kleinian', a method of presenting structure via privileged charts must identify the structure via its invariance under a group or pseudogroup defined by the charts. We do not agree that Lorentz affine charts cannot be employed in such a manner. The rest of this section explains how they can be so used, but the basic idea is straightforward. Rather than considering the totality of the charts in order to define a single pseudogroup, one instead considers the groups and related structures defined by the charts centred on each point of $M$ considered separately.

Suppose that, for some particular relativistic spacetime $(M,g)$, one is presented with all and only the adapted charts corresponding to one of the conditions described in the previous section. Call this set $C$. Two questions naturally arise. First, provided only with this information, can one identify the condition satisfied by all the charts in $C$? Second, can one further recover the structure of $(M,g)$ from the charts and their transition functions?\footnote{Note that, because there is a set of adapted charts for every point of $M$, the collection of all such sets is an atlas for $M$ and therefore we are always able to recover $M$'s differential structure.}

The answer to the first question is `yes'. By design, each chart is adapted to the geometrical structure around the point at the origin of the chart. There is therefore a crucial, and easily defined, partition on $C$. The elements of the partition are indexed by the points of $M$, with two charts belonging to the same element just in case they assign $(0,0,0,0)$ to the same point. In terms of transition functions, $\phi, \psi \in C$ belong to the same element if and only if $\psi \circ \phi^{-1} \colon (0,0,0,0) \mapsto (0,0,0,0)$.

In order to see which coordinate condition the charts implement, one need only attend to the transition functions between charts in the same partition element. For each condition, these transition functions will satisfy a characteristic constraint. In the case of normal and Lorentz normal charts, we have already said what these are in \S\ref{s:normal} and \S\ref{LNCs}%
. The constraints associated with local Lorentz charts will be described shortly. The constraints associated with locally affine charts are described in Appendix~\ref{appendixB}.

Having identified the nature of the charts, can one recover the structure of $(M,g)$? The answer depends on which condition the charts implement. From, e.g., the local conformal charts one can at most hope to recover the conformal structure of $(M,g)$. But this will not allow one to determine all the structure of $(M,g)$ because non-isometric spacetimes can share their conformal structure. As Barrett and Manchak note, the Lorentz affine charts do allow one to recover all structure. But they do so because the local Lorentz charts already suffice to recover all structure, as we shall now spell out.

First, note that a chart $\phi$ defined on a neighbourhood of $p$ defines a corresponding coordinate system on $T_pM$, the tangent space at $p$. The tangent vectors $\partial_{\mu}|_{p}$ to the coordinate curves at $p$ provide a basis, the \textit{coordinate basis}, for $T_pM$. Recall that a basis for a $4$-dimensional vector space, $V$, is just a way of specifying a bijection from $V$ to $\mathbb{R}^4$ that qualifies as an adapted coordinate system in the sense of \S\ref{pc}. It is an invertible linear map from $V$ to $\mathbb{R}^4$ considered as a vector space. To say that $T_pM \ni X = X^{\mu}\partial_{\mu}$ is just to say that the coordinates of $X$ in the induced adapted coordinate system are $(X^0, X^1, X^2, X^3)$.

If $p$ is in the domain of chart $\phi$, let $T \phi_p \colon T_pM \to \mathbb{R}^4$ be the corresponding adapted induced coordinate system for $T_pM$. This slight abuse of notation can be justified as follows. Associated with the map $\phi \colon U_p \to  \mathbb{R}^4$ there is an induced map $T\phi \colon TU_p \to  T\mathbb{R}^4$, call it the tangent map,  from the tangent bundle over $U_p$ to the tangent bundle over $\mathbb{R}^4$. For any point $q \in U_p$, this map gives us an isomorphism, $T\phi_q$%
, from $T_qM$, the tangent space at $q$, to $T_{\phi(q)}\mathbb{R}^4$, the tangent space at $\phi(q)$. If we identify the latter with $\mathbb{R}^4$ in the canonical way, this is just our adapted coordinate system for $T_qM$.

Let $C^{g}_p$ be the set of local Lorentz charts of $(M,g)$ at $p$. Recall that these are all and only the charts that send $p$ to the origin of $\mathbb{R}^4$ and in which $g_{\mu \nu}|_p = \eta_{\mu \nu}$. This means that, as \textcite[p.~16]{BM3} note, there is an easy way to recover $g_{ab}$ from the set of all local Lorentz charts $\{C^g_p : p \in M \}$. $g_{ab}$ is just an assignment of a Minkowski inner product to the tangent space of every point of $M$. Each and every member of $C^g_p$ can be used to \textit{explicitly} define the relevant inner product on the tangent space at $p$. Let $\phi \in C^g_p$ and let $X^{\mu}$ be the components of $X \in T_pM$ according to $T\phi_p$. Then for any $X, Y \in T_pM$, $g(X,Y) := -X^0Y^0 + X^1Y^1 + X^2Y^2 + X^3Y^3$. Repeating this definition for every point in $M$ gives us the inner product on every tangent space and therefore recovers $g_{ab}$.

Associated with the explicit definition of this structure there is a corresponding Kleinian characterisation of this structure. The explicit definition effectively uses $T\phi_p$ to pull back to $T_pM$ (one of) the natural definition(s) of the Minkowski inner product on $\mathbb{R}^4$. The transformation group on $\mathbb{R}^4$ that preserves this inner product is just the Lorentz group, $O(1,3)$. Hence if $d$ is a diffeomorphism on $\mathbb{R}^4$, the coordinate system $d \circ T\phi_p$ will define the same inner product on $T_pM$ if and only if $d \in O(1,3)$. In other words, in line with the approach of \textcite{wallace-coord}, we can think of the set $TC^g_p := \{T\phi_p : \phi \in C^g_p \}$ as a set of adapted charts on $T_pM$, where any two $\phi, \phi' \in C^g_p$ will satisfy the group-theoretic constraint $T\phi'_p \circ (T\phi_p)^{-1} \in O(1,3)$. \label{fn-ortho}%
If $x^\mu$ are the coordinate functions of $\phi$, and $y^\mu$ those of $\phi'$  (with $y^\mu$ understood as functions of $x^\mu$) we can rewrite this condition as $\frac{\partial y^{\nu}}{\partial x^{\mu}}|_p\in O(1,3)$.

Alternatively, in the spirit of Barrett and Manchak's version of the Kleinian approach, we can think of the set $TC^g_p$ as symmetry-specifying charts and characterise the inner product on $T_pM$ as exactly that structure left invariant by the transformation group on $T_pM$ defined via $\Gamma = \{(T\phi_p)^{-1} \circ T\phi'_p: \phi, \phi' \in C^g_p \}$.

We take the foregoing to demonstrate clearly how an arbitrary Lorentzian metric can be recovered just from the full set of its local Lorentz charts in a manner clearly in the spirit of the Kleinian approach. Notably, no further restriction to locally affine charts is required. For those to whom this might still seem magical, we offer a couple of additional observations.

First, it is worth stressing that, in being provided with the full set of local Lorentz charts for some metric, one is provided not only with (i) the transition functions between charts in the same point-indexed equivalence classes but also (ii) the transition functions between charts from different equivalence classes whose domains nevertheless intersect. Only (i) is used in ascribing an inner product to the tangent space of each point %
but (ii) is crucial in distinguishing between different Lorentzian geometries on the manifold. With only (i) all we have is a collection of manifolds diffeomorphic to $\mathbb{R}^4$, each containing a single point on the tangent space of which a Minkowski inner product is defined. To recover both the single manifold and its unique metric field, we need to know how these manifolds knit together.

Second, one might consider how the local Lorentz charts allow one fairly directly to fix the metric via an identification of a privileged class of curves. For each curve, simply ask whether, at each point on the curve, there is at least one local Lorentz chart such that the curve's tangent vector at that point has coordinates $(1,0,0,0)$. This condition is satisfied by all and only everywhere timelike curves parameterised by proper time. And the class of such curves fixes the metric.\footnote{See \parencite[p.~125]{Malament2012}. The class of smooth timelike curves gives the conformal structure of spacetime; the preferred parameterisation of those curves yields a volume element which, in conjunction with the conformal structure, suffices to fix the metric.}

Up until this point we have assumed an antecedently-given relativistic spacetime $(M,g)$. We have reviewed how certain charts, in particular the local Lorentz charts, can be adapted to its structure and described how that structure can in turn be recovered from those charts. We now consider a more ambitious project. Can one characterise in coordinate-based terms properties that are sufficient for a collection of charts to define a relativistic spacetime structure?

The shape of a positive answer to these questions is already evident from our description of the recovery process from local Lorentz charts. The question effectively becomes: what are the necessary and sufficient conditions for a set of charts on $M$ to constitute the set of all the local Lorentz charts for some metric $g$ defined on $M$?

We start by defining $M$'s differential and topological structure via a choice of maximal atlas $\mathcal{A}$ for $M$. Next, consider the sub-atlas $\mathcal{A}_0$ defined via the condition that its charts include the origin of $\mathbb{R}^4$ in their range. That is, $\mathcal{A}_0 = \{ \phi : \phi \in \mathcal{A} \text{ and for some } p \in M, \phi(p) = (0,0,0,0) \}$. We now define a double partition of $\mathcal{A}_0$. At the first level, charts are grouped according to which point is at the origin of the charts. At the second level, each element of the first partition is further partitioned according to the inner product induced on the tangent space of the point at the origin of the charts.

In more detail, the elements of the first partition are indexed by the points of $M$. We denote the partition as $\{C_q : q\in M \}$, where a chart is in the equivalence class $C_q$ just in case $q$ is at the origin of that chart: $C_q := \{ \phi : \phi \in \mathcal{A}_0 \text{ and } \phi(q) = (0,0,0,0) \}$. %

We now consider the further partition of the elements of $\{C_q : q\in M \}$. Consider a specific element $C_p$ of the partition and consider the relation defined on its charts via
\begin{equation}\label{eq:sim}
    \phi\sim \bar\phi \quad \text{iff}\quad (T\bar\phi_p) \circ (T\phi_p)^{-1} \in O(1,3)\subset GL(4),
\end{equation}
with $\phi, \bar\phi\in C_p$. This relation is obviously an equivalence relation and so defines a partition on $C_p$.

Any chart $\phi$ in $C_p$ defines a Minkowski inner product $\eta_{\phi}$ on $T_pM$ via the condition: if $X,Y \in T_pM$ and the components of $X$ and $Y$ in $T\phi_p$ are $X^{\mu}$ and $Y^{\mu}$, then $\eta_{\phi}(X,Y) = -X^0Y^0 + X^1Y^1 + X^2Y^2 + X^3Y^3$. As described above (see also Appendix~\ref{appendix}), any two charts in $C_p$ define the same inner product if and only if they stand in the equivalence relation defined by condition~\eqref{eq:sim}. We will index the elements of the second-level partition by the inner product that their charts define. That is, we will denote an element of this partition as $C^{\eta(p)}_p$ where $\eta(p)$ is the inner product defined on $T_pM$ by any chart in $C^{\eta(p)}_p$.\footnote{Writing `$\eta(p)$' rather than just `$\eta$' might seem to clutter notation, but it will aid clarity in Appendix~\ref{appendix}.}

The upshot is that we can define a metric on $M$ by picking, for each point $q \in M$, an element $C^{\eta(q)}_q$ of the second-level partition of the corresponding element $C_q$ of the first-level partition. In other words, a family $\{C^{\eta(q)}_q : q\in M \}$, where $\eta(q)$ is also dependent on $q$, corresponds to a metric over $M$. Moreover, each distinct such family corresponds to a different metric on $M$, and each metric on $M$ corresponds to exactly one such family.

In general, a metric defined in this way will not be smooth or even continuous. For each $p \in M$, our choice of element in the partition of $C_p$ was entirely unconstrained by our choices of elements in the partitions of $C_q, q \neq p$. A metric that is not continuous is of very limited use.\footnote{For example, one cannot use it to define arc lengths along curves.} It is thus natural to ask what further coordinate conditions might be imposed to ensure that a metric defined in this way is smooth or at least continuous. We address this question in Appendix~\ref{appendix}. In Appendix~\ref{appendixB}, we provide a parallel implicit definition of an affine connection in terms of privileged charts (but leave smoothness as an exercise to the reader).

\section{Close}\label{close}

The connection between Lorentzian spacetimes and charts meeting the condition just described (i.e., a family of charts of the form $\{C^{\eta(q)}_q : q\in M \}$) is both a uniqueness and an existence result. A set of charts meeting the condition always corresponds to the set of local Lorentz charts for a unique Lorentzian spacetime.\footnote{``Proposition 6'' in \textcite{BM3} is a corresponding uniqueness claim for Lorentz affine charts. We have highlighted that the restriction of the Lorentz charts to locally affine charts does no work.} That it is an existence result is worth stressing. We have shown that if one begins with a set of local Lorentz(-like) charts without having antecedently specified a Lorentzian metric, one can always \emph{construct} such a metric.

It is natural to ask whether similar results could be achieved via a different set of charts (a set that is not merely a proper subset of the local Lorentz charts). There is at least one natural candidate. For a given spacetime one could take the set of all of its local conformal charts that are also locally projective charts. This set will not be a subset of the local Lorentz charts. From it  one will be able to recover both conformal and projective structure. One can then appeal to the uniqueness result famously demonstrated by \textcite{Weyl1921Translation}, that a Lorentzian metric $g$ can be recovered, up to homothety, from its associated projective and conformal structure.\footnote{For a modern presentation of this result, see \parencite[ch.~2]{Malament2012}; for further recent discussion, see \parencite{AdlamLinnemannRead}.}

That gives us a uniqueness result---the Lorentzian geometry that can be recovered from the conformal projective charts of a spacetime is unique---but what about a corresponding existence result? There is a precedent in the literature for such existence results: famously, \textcite{EPS} purported to demonstrate that a Lorentzian metric can be fixed by projective and conformal structures, together with a `compatibility' condition and other auxiliary assumptions. It would be interesting to translate this approach into a chart-based construction and, in particular, to consider what it takes, in chart-based terms, for the conformal structure specified by one set of charts and the projective structure specified by another to be  compatible.\footnote{Again, for recent discussion of the result of \textcite{EPS}, see \parencite{AdlamLinnemannRead}.}

Stepping back somewhat, we see some quite significant philosophical applications of this work. The programme of `regularity relationalism' due to \textcite{huggett2006regularity} seeks to offer a Humean reduction of some elements of spacetime structure. As per \textcite{Lewis1973-LEWC-2}, dynamical laws are reduced to the simplest and strongest codifications of the local matters of fact constituting the Humean mosaic. Huggett's innovation is to  reduce in addition elements of spacetime structure to the structure preserved by the coordinate transformations relating preferred coordinate expressions of those laws. It is therefore a way of making precise (and providing a metaphysical underpinning for) a recurrent idea that physical geometry might be given a Kleinian reduction in terms of the symmetries of the dynamical laws.\footnote{See, e.g., \parencite{anandan1980hypotheses,brown2005space}. \textcite[\S6.3.2]{pooley2013substantivalist} argues that Brown's view should be interpreted in this manner. The framework is explored further by \textcite{stevens20reg}.}

This approach, however, applies most naturally to global dynamical symmetries and thus to globally homogeneous spaces. There remains a challenge regarding whether (if at all) it can be applied to theories such as general relativity.\footnote{See, e.g.,\ \parencite{Dewar2020-DEWGC-3}.} The machinery developed in this article should provide much of the technical wherewithal to make good on this project. The idea, quite naturally, would be to regard some subset of the local Lorentz charts as picked out via codifications of \emph{local} goings-in a Humean mosaic, and then to use an existence result of the kind mentioned above in order to obtain a Lorentzian metric. Evidently, though, this is but a sketch; we will make good on that particular project elsewhere.

\appendix

\section{Defining a Smooth Metric via Privileged Charts}\label{appendix}

As described in \S\ref{recovering} above, a family of charts $C^{\eta} = \{C^{\eta(q)}_q : q\in M \}$, where $\eta(q)$ is also dependent on $q$, corresponds to a metric over $M$, but this metric is not necessarily smooth. In order to obtain smoothness directly from conditions on coordinate charts, we need to impose further constraints.

Picking out a point $p \in M$ and a chart $\phi_p \in C^{\eta(p)}_p \in C^{\eta}$ at that point, pick another $p' \neq p$ in the domain of $\phi_p$  and consider $\phi_{p'}\in C_{p'}^{\eta(p')} \in C^{\eta}$. 
 Let $x^\mu$ be  coordinates for $\phi_p$ and $x^{\mu'}$ be coordinates for $\phi_{p'}$.
 The latter chart gives vectors tangent to its coordinate curves at its origin: $\{\partial_{\mu'}|_{p'}\in T_{p'}M\}$, and this basis will have coordinate components in $\phi_p$, which we write as a $4\times 4$ component matrix: $M\indices{^{\mu}_{\mu'}}(p')$, which we index with the point $p'$, as this will be useful for the construction that follows. 
 We then of course have 
 \be M\indices{^{\mu}_{\mu'}}(p')=\frac{\partial x^{\mu}}{\partial  x^{\mu'}}(p\rq{})\in GL(4).
 \ee 

In what follows, we will have to track many different coordinate systems, so it pays to write this matrix abstractly without explicit coordinates as the invertible linear transformation: 
\begin{equation}\label{eq:M}
M(p'):= ((T\phi_p)_{p'})^{-1}\circ (T \phi_{p'})_{p'}
 \end{equation}
Note that, in an expression such as $(T\phi_{p})_{p'}$, the first subscript means that the chart is adapted to the point $p$, i.e.\ is an element of $C_{p}$, and the second tells us that we are considering the restriction of the tangent map of that chart to the point $p'$. %

Now to our main definition, of smoothness of a family of charts:
 
 \begin{defi}[Smoothness of a family $\{C^{\eta(q)}_q,\, q\in M\}$] Given a smooth curve wholly within the domain of a chart $\phi_p\in C_p^{\eta(p)}$,  i.e.\ an embedding $\gamma:I\rightarrow {(\phi_p)}^{-1}(\mathbb{R}^4)\subset M$, such that $\gamma(0)=p$, a given 1-parameter family $C^{\eta(t)}_{\gamma(t)}$ is smooth along $\gamma$  iff for each $t$ there exists a $ \phi_{\gamma(t)}\in C^{\eta(t)}_{\gamma(t)}$ with $ C^{\eta(0)}_{\gamma(0)}=C_p^{\eta(p)}$ and $\phi_{\gamma(0)}=\phi_{p}$ such that the corresponding 1-parameter family of matrices $M(\gamma(t))\in GL(4)$ is smooth, where:
 \begin{equation}\label{eq:M_t}
     M(\gamma(t)):=((T\phi_p)_{\gamma(t)})^{-1}\circ (T\phi_{\gamma(t)})_{\gamma(t)},
 \end{equation}
 such that $M(\gamma(0))=\text{\em Id}$.  \label{def:smooth}
 \end{defi}

 We would like to use this notion to define smoothness for metrics corresponding to families of inner products as described by $\{C^{\eta(q)}_q : q\in M\}$.
\begin{theo}\label{thm:1}
   A given family of $\{C^{\eta(q)}_q : q\in M\}$ corresponds to a smooth metric iff the restriction of this family to any smooth curve $\gamma$ is smooth according to Definition \ref{def:smooth}. 
\end{theo}

\noindent \emph{Proof:} Take $p, p'\in M$. Now  define $\langle X, Y\rangle_{p'}:=  -X^0Y^0 + X^1Y^1 + X^2Y^2 + X^3Y^3$,  where $X(p')=X^{\mu'}(p')\partial_{\mu'}|_{p'}$ and $Y(p')=Y^{\mu'}(p')\partial_{\mu'}|_{p'}$ for $x^{\mu'}$ the coordinates of $\phi_{p'} \in C^{\eta}_{p'}$ and, as before, $\partial_{\mu'}|_{p'}\in T_{p'}M$. It is clear from \eqref{eq:sim} that any such metric depends only on the family $\{C^{\eta(p')}_{p'} : p' \in M\}$. The challenge now is to represent the metric at each point not according to a chart adapted to that point, but rather according to one single chart, for which we can show smoothness. 

In the chart $\phi_p \in C^{\eta(p)}_{p}$, we rewrite $\partial_{\mu'}|_{p'}=M\indices{^{\nu}_{\mu'}} \partial_{\nu}|_{p'}$, where, as before, $\partial_{\nu}|_{p'}$ now refers to the chart $\phi_p$. We obtain
\begin{equation}\label{eq:etap'}
  \langle \partial_{\mu'} , \partial_{\nu'}\rangle|_{p'}:=  \eta_{\mu'\nu'}|_{p'}=\eta_{\mu'\nu'}=M\indices{^{\mu}_{\mu'}} M\indices{^\nu_{\nu'}}\langle \partial_{\mu} , \partial_{\nu}\rangle|_{p'}.
\end{equation}
We get the inner product at $p'$ for a basis that is not adapted to $p'$ as $\eta_{\mu\nu}|_{p'}:=\langle \partial_{\mu}, \partial_{\nu}\rangle_{p'}$ in  \eqref{eq:etap'}. Thus,
\begin{equation}\label{eq:relation_M}
    \eta^{p'}_{\mu\nu}:=\langle \partial_{\mu}, \partial_{\nu}\rangle_{p'}=(M^{-1})\indices{^{\mu'}_{\mu}}(M^{-1})\indices{^{\nu'}_{\nu}}\eta_{\mu'\nu'},
\end{equation}
which gives the inner product in a coordinate chart adapted to $p$ in terms of an inner product in a coordinate chart adapted to $p'$.
Setting $p'=\gamma(t)$, the metric in \eqref{eq:relation_M} is smooth along a curve iff the matrices $M$ are smooth along that curve.

To see that smoothness of the metric in this chart-dependent definition  depends only on membership of the family $\{C^{\eta(q)}_q : q\in M\}$,  suppose we have a second family of charts $\{ \tilde\phi_{q}\in C^{\eta(q)}_{q}: q\in M\}$.
By definition \eqref{eq:sim},
\begin{equation}
    ((T\tilde\phi_{p'})_{p'})^{-1}\circ ((T\phi_{p'}))_{p'}=\Lambda(p')\in O(1,3), 
\end{equation}
understood in terms of the $\mathbb{R}^4$ coordinates, i.e.\ such that \begin{equation}\label{Lambda_eta}\Lambda\indices{^\mu_{\tilde\mu}}\Lambda\indices{ ^\nu_{\tilde\nu}}\eta_{\mu\nu}=\eta_{\tilde\mu\tilde\nu}.\end{equation}

  For $p'$ within the domain of both $\tilde\phi_p$ and $\phi_p$, we can write: 
\begin{align*}
    \tilde M(p')&:=((T\tilde\phi_p)_{p'})^{-1}\circ (T\tilde\phi_{p'})_{p'}\\
   &=((T\tilde\phi_p)^{-1}\circ (T\phi_p))\circ ((T\phi_p)^{-1}\circ (T\phi_{p'}))\circ ((T\phi_{p'})^{-1}\circ (T\tilde\phi_{p'})_{p'})\\
   &=\Lambda(p) M(p')\Lambda(p')^{-1}.
\end{align*}
(In the middle line, some of the outer $p'$ subscripts have been suppressed for legibility.%
) It follows from this and \eqref{Lambda_eta}, that for the tilded version---i.e. adding tildes to all the super and subscripts, including the primed ones---of  equation  \eqref{eq:relation_M}, which defines the metric at all points in the domain of a chart, $\Lambda(p')$ acts on the inner product in a coordinate chart adapted to $p'$, leaving that inner product invariant. Thus
\begin{align*}
    \eta^{p'}_{\tilde\mu\tilde\nu}&=(\tilde M^{-1})\indices{^{\tilde\mu'}_{\tilde\mu}}(p')(\tilde M^{-1})\indices{^{\tilde\nu'}_{\tilde\nu}}(p')\eta_{\tilde\mu'\tilde\nu'}\\&=\Lambda(p)\indices{^{\mu}_{\tilde\mu}}\Lambda(p)\indices{^{\nu}_{\tilde\nu}}(M^{-1})\indices{^{\mu'}_{\mu}}(p')(M^{-1})\indices{^{\nu'}_{\nu}}(p')\Lambda(p')\indices{^{\tilde\mu'}_{\mu'}}\Lambda(p')\indices{^{\tilde\nu'}_{\nu'}}\eta_{\tilde\mu'\tilde\nu'}
    \\
    &=\Lambda(p)\indices{^{\mu}_{\tilde\mu}}\Lambda(p)\indices{^{\nu}_{\tilde\nu}}(M^{-1})\indices{^{\mu'}_{\mu}}(p')(M^{-1})\indices{^{\nu'}_{\nu}}(p')\eta_{\mu'\nu'},
\end{align*}
where we inserted dependence on $p'$ for clarity. Thus dependence on $p'$ amounts only to that already contained in $M$ (recall that $\eta_{\mu'\nu'}$ is just the diagonal matrix with Minkowski signature). Moreover, it is easy to verify that, since $M(p)=\text{Id}$ (and $\Lambda(p)\in O(1,3)$), we still get 
\begin{equation}
\eta^{p}_{\tilde\mu\tilde\nu}=\Lambda(p)\indices{^{\mu}_{\tilde\mu}}\Lambda(p)\indices{^{\nu}_{\tilde\nu}}\eta_{\mu\nu}=\eta_{\tilde\mu\tilde\nu},
\end{equation}
as required by consistency.

So, as far as the metric dependence on $p'$ goes, we can still use the same matrices $M$, but with tilded indices, i.e.\ indices understood as elements of the tilded family of charts.%

This shows that the metric in \eqref{eq:relation_M}, written with respect to a chart, is smooth iff the matrices $M$ in \eqref{eq:M}  
are smooth, for any family of charts adapted to $\{C^{\eta(q)}_q : q\in M\}$. $\square$

\section{Defining a Connection via Privileged Charts}\label{appendixB}

The methods applied to local Lorentz charts in \S\ref{recovering} can be applied to charts adapted to other kinds of structure. Here we briefly examine the case of affine structure.

\subsection{Recovery}

Let $(M,\nabla)$ be a manifold equipped with an affine connection. Let 
$\phi$ and $\bar\phi$ be two charts defined on neighbourhoods of $p \in M$ with $p$ at the origin. Let $x^\mu$ and $\bar x^\mu$ be their corresponding coordinate functions. Recall from \S\ref{affine} that a chart is \emph{locally affine} at $p \in M$ if the connection coefficients vanish at $p$, the origin of the chart.  Suppose that $\phi$ is a locally affine chart at $p$. Recall that the transformation for the connection coefficients is given by:
\be\label{eq:Christ} {\bar{\Gamma}^i}_{kl} =
  \frac{\partial \bar{x}^i}{\partial x^m}\,
  \frac{\partial x^n}{\partial \bar{x}^k}\,
  \frac{\partial x^p}{\partial \bar{x}^l}\,
  {\Gamma^m}_{np}
  +
  \frac{\partial^2 x^m}{\partial \bar{x}^k \partial \bar{x}^l}\,
 \frac{\partial \bar{x}^i}{\partial x^m}
\ee
It follows that $\bar \phi$ will also be locally affine at $p$ %
iff:
\be\label{eq:proj}
\left. \frac{\partial^2 \bar x^\mu}{\partial x^{\nu}\partial x^\rho}\right|_{p}=0.
\ee

When recovering an inner product at a point from local Lorentz charts, we provided an explicit definition in terms of an arbitrarily selected chart centred on that point. We do the same in this case. We define the derivative operator $\nabla^{\phi}$ associated with a chart $\phi$ via:
\be\label{defcon}
\nabla^{\phi}_X Y:=(X^\mu \partial_\mu Y^\nu) \partial_\nu.
\ee
Equation~\eqref{eq:proj} is then just the condition that actions of the derivative operators associated with charts $\phi$ and $\bar{\phi}$ agree at $p$. Since a connection is defined by its action at every point, we recover $\nabla$ from the full set of locally affine charts as the connection the action of which agrees at each point $p \in M$ with the actions of the connections defined by the locally affine charts at $p$.

\subsection{Construction}

Next, again following the approach of \S\ref{recovering}, we ask what conditions can be imposed on a set of charts in order for those charts to define a connection. We start with the same first-level partition $\{C_q : q\in M\}$ of the atlas $\mathcal{A}_0$. But now, instead of defining a second-level partition via condition~\eqref{eq:sim}, we invoke, for each $p \in M$, equation~\eqref{eq:proj}. It is straightforward to verify that this defines an equivalence relation on charts and thus also defines a partition on each $C_p$%
. Consider now a specific element, which we label $C^{\nabla(p)}_p$, of this second-level partition of $C_p$. We can define affine connections associated with each chart in $C^{\nabla(p)}_p$ via equation \eqref{defcon}. As we have just seen, their actions will all agree at $p$. Thus a choice of element from the second-level partition for each $p \in M$ defines a family of charts $\{C^{\nabla(p)}_p : p \in M\}$ that determines a unique covariant derivative operator at each point $p \in M$. Such a family thus defines an affine connection on $M$.

As before, nothing guarantees the smoothness of such a family. As a result, the connection that it defines might lack some desirable features.\footnote{For example, there might be no smooth curves in $M$ the tangent vectors of which are everywhere auto-parallel according to the defined connection.} In Appendix~\ref{appendix} we described a condition on a family $\{C^{\eta(p)}_p : p \in M\}$ that guaranteed the smoothness of the metric that it defines. One might wish to define an analogous condition (perhaps in terms of the charts to be described in the next section) on the family $\{C^{\nabla(p)}_p : p \in M\}$ to ensure the smoothness of the connection that it defines. We leave this as an exercise for the interested reader.

 \subsection{The Kleinian Perspective}
 
There is a natural group-theoretic characterisation of the condition relating the coordinate transformations between two local Lorentz charts, namely: $\frac{\partial y^{\nu}}{\partial x^{\mu}}|_p \in O(1,3)$. Associating the coordinate transformations between charts with a group played a vital role in connecting the charts to the symmetry group of the structure defined by the charts. In other words, it was vital to a Kleinian interpretation of the charts. Our statement above of the condition relating locally affine charts, namely equation \eqref{eq:proj}, was not group-theoretic. We now show how a group-theoretic characterisation can be given.

Whereas a metric field on $M$ involves assigning structure to the tangent space at each point of $M$, an affine connection can be thought of as assigning structure to the tangent space at every point of the tangent bundle $TM$ of $M$; i.e. it concerns the `double' tangent bundle. We first describe this structure before relating its symmetry group to coordinate transformations.\footnote{Indeed, given any vector bundle $E$ over $M$, we can define $TE$ as another vector bundle over $M$; see \parencite[\S8.12]{Michor2008}.}

For ease of comparison with our treatment of local Lorentz charts, let $M$ be a $4$-dimensional manifold. $TM$ is then $8$-dimensional, as is the tangent space $T_{(p,v)}(TM)$ at each point $(p,v) \in TM$, $p \in M, v \in T_pM$. Elements in this space are tangent to  the doublet of curves, $(\gamma(t), X(t))\in TM$, $\gamma(t) \in M, X(t)\in T_{\gamma(t)}M$ that go through $(p,v)$ at $t=0$. In virtue of the construction of $TM$, each space $T_{(p,v)}(TM)$ has a privileged $4$-dimensional subspace $V_{(p,v)}$, the \emph{vertical subspace}, consisting of all and only vectors that are tangent to $T_pM$ considered as a submanifold of $T_{(p,v)}(TM)$; this is generated by curves $(\gamma(t), X(t))$ for which $\gamma(t)=p$ for every $t$.\footnote{One can see the vertical subspace as the pullback of the projection $T(TM)\rightarrow TM$ over the zero section.} A \emph{horizontal subspace}, $H_{(p,v)}$, is a choice of $4$-dimensional subspace such that $H_{(p,v)} \oplus V_{(p,v)} = T_{(p,v)}(TM)$.\footnote{In other words, it corresponds to a choice of projection $\hat V: T(TM)\rightarrow V$ \parencite[cf.][\S17.3]{Michor2008}. Here we are extending the definition of $V_{(p,v)}$ to every $(p,v)$: this forms a sub-bundle $V\subset T(TM)$. Then a projection $\hat V: T(TM)\rightarrow V$ is such that $\hat V\circ \hat V=\hat V$ and $\text{Im}(\hat V)= V$.} In these terms, an affine connection on $M$ corresponds to a choice of horizontal subspace for each $T_{(p,v)}(TM)$.\footnote{Heuristically, this is because a connection gives us a way to compare vectors in the tangent spaces at `infinitesimally separated points'. The choice of a horizontal subspace makes this precise as follows. Curves in $TM$ whose tangent vectors lie everywhere within a vertical subspace `live' inside the tangent space of a point, and so cannot `thread' the tangent spaces belonging to neighbouring points. The complementary horizontal subspaces provide a threading of the vector spaces; for each path in $M$ they provide a linear map between the tangent spaces of the points on the path. 
\label{ftnt:hor}}
Given a choice of horizontal subspace, we can write elements of $T(TM)$ as $(p, v, Y, Z)$, with 
\begin{equation}
\label{eq:HV_dec} (p, v, Y, 0)\in H_{(p,v)}\, ; \quad(p, v, 0, Z)\in V_{(p,v)}.
\end{equation}

In our characterisation of a metric in terms of local Lorentz charts, we exploited the fact that, for any given chart $\phi$ defined on $U_p \in M$, there are natural coordinatisations of the tangent spaces $T_qM$, $q \in U_p$, i.e., those given by the coordinate basis at each point (comprised of the tangent vectors to the coordinate curves at each point). In other words, each chart is associated with a coordinatisation of the tangent bundle $TU_p$ over its domain. In turn, this coordinatisation of $TU_p$ is associated in exactly the same way with a coordinatisation of $T(TU_p)$, which is what we will use here. 

In slightly more detail, each chart $\phi$ provides a natural basis for $T_{(p,v)}(TM)$ with $p \in \text{Dom}(\phi)$ and $v \in T_p(M)$. At any such point $(p,v)$, $4$ of these $8$ basis vectors will be tangent to $T_p(M)$. These span the vertical subspace of $T_{(p,v)}(TM)$. The other $4$ basis vectors define the horizontal subspace that define the chart's connection at that point. Any transformation of charts that doesn't mix coordinates of the vertical subspaces with those of the adapted horizontal subspaces will therefore preserve the horizontal subspace. 

More specifically, suppose $x^\mu$ are the coordinate functions of $\phi$, then the coordinates $v^\mu$ of a tangent vector $v$ are given by $v^\mu \partial_\mu$. In other words, we map an element $(p,v) \in TM$ to its coordinates $(x^\mu, v^\mu)\in \mathbb{R}^4\times \mathbb{R}^4$. These are coordinates for $TM$, seen as manifold, and we can repeat the process for $T(TM)$, obtaining a map: 
\begin{align}
   TT\phi:T(TU_p) &\rightarrow \mathbb{R}^4\times \mathbb{R}^4\times \mathbb{R}^4\times \mathbb{R}^4\nonumber \\
   \zeta &\mapsto (x^\mu, v^\mu, Y^\mu, Z^\mu)\label{TTM},    
\end{align}
which could, for mnemonic reasons, also be written as: 
\begin{equation}
    (x^\mu, \partial x^\mu, \delta x^\mu, \delta \partial  x^\mu),
\end{equation}
where $\partial$ if the tangent map going from $p\in M$ to $T_pM$, and $\delta$ is the tangent map going from $(p,v)\in TM$ to $T_{(p,v)}(TM)$. 
Vertical vectors can be locally written as those with $\delta x^\mu=0$.\footnote{Cf.\ footnote \ref{ftnt:hor}.} 
We now consider how a change of coordinates affects the decomposition \eqref{TTM}, i.e.\ we
 consider transformations on $T_{(p,v)}(TM)$ induced by a transformation between two charts whose domains both contain $p$. To shorten notation, let us call the composition $\psi:=\phi^{-1}\circ \bar\phi:\mathbb{R}^4\rightarrow \mathbb{R}^4$. So, in coordinates, we have:
 \begin{align}
     \psi' &:=\frac{\partial \bar x^\mu}{\partial x^\nu}: \mathbb{R}^4\rightarrow \mathbb{R}^4,\\
     \psi'' &:= \frac{\partial^2 \bar x^\mu}{\partial x^{\nu}\partial x^\rho}:\mathbb{R}^4\times \mathbb{R}^4\rightarrow \mathbb{R}^4,
 \end{align}
 which acts on components written in unbarred indices. 
Now we write the induced transition for the double tangent bundle (omitting indices for now), taking the decomposition according to $\phi$ to that according to $\bar \phi$: 
 \begin{equation}
     (TT\psi)(x,v,Y,Z) = (\psi(x),\psi'(x)(v),\psi'(x)(Y), \psi''(x)(v,Y) + \psi'(x)(Z)).\label{eq:trans}
 \end{equation}
Clearly, a transition function will preserve the vertical space: if a vector in $T_{(p,v)}(TM)$ has no horizontal component according to $\phi$, i.e. $Y=0$ as per \eqref{eq:HV_dec}, then $\psi'(x)(Y)=0$, meaning it has no horizontal component according to $\bar \phi$.\footnote{Which is as it should be, since the vertical space is defined canonically: if a vector in $T_{(p,v)}(TM)$ has no horizontal part, so that $Y=0$, it is purely vertical according to every decomposition into vertical and horizontal.} %
But generally this transition will not preserve the horizontal complement to the vertical space. That is, given one vector and two charts, \textit{the horizontal parts} of that vector according to the two charts are related by an element of $GL(4)$, but the entire vector may be horizontal in one chart and not in the other. Mathematically, while $Z=0$, $\bar Z$ can be non-zero; the vector which lacked a vertical component according to $\phi$ acquires one according to $\bar \phi$. The  term in the transition function \eqref{eq:trans} that mixes vertical and horizontal vectors is%
 \begin{equation}
     \psi''(x)(v, Y)=\frac{\partial^2 \bar x^\mu}{\partial x^{\nu}\partial x^\rho} v^\nu Y^\rho.
 \end{equation}
Therefore, when \eqref{eq:proj} is satisfied $\psi''$ vanishes and the horizontal subspace is preserved. In this case, it follows that $\bar Z^\mu=\Lambda\indices{^\mu_\nu} Z^\nu$ for $\Lambda\in GL(4)$. So we could define adapted charts at a point as those whose both horizontal and vertical components are individually related by $GL(4)$.  %

In sum, the transformation group acting on $T_{(p,v)}(TM)$ for some $v \in T_pM$ associated with transformations between the set of locally affine charts at $p$ has the form $GL(4) \times GL(4)$ (rather than a subgroup of $GL(8)$ that only preserves the vertical subspace). The horizontal subspace defined via the criterion of invariance under this group is just the subspace associated with the connection, and the set of the subspaces picked out in this way for each $v \in T_pM$, defines the action of the connection at $p$. We can therefore regard the privileged charts associated with a connection---the locally affine charts---as defining their connection via a natural generalisation of the Kleinian method.%

\printbibliography

\end{document}